# Analysis of interdiffusion between $SmFeAsO_{0.92}F_{0.08}$ and metals for *ex situ* fabrication of superconducting wire

M Fujioka[1], M Matoba[1], T Ozaki[2], Y Takano[2,3], H Kumakura[2,3] and Y Kamihara[1,3]

[1]Department of Applied Physics and Physico-Informatics, Faculty of Science and Technology, Keio University, 3-14-1 Hiyoshi, Yokohama 223-8522, Japan

[2]National Institute for Materials Science, 1-2-1 Sengen, Tsukuba 305-0047, Japan

[3]JST, TRIP, Sanban-cho bldg. 5, Sanban-cho, Chiyoda, Tokyo 102-0075, Japan

E-mail: fujioka-masaya@a6.keio.jp
TEL: +81 45-563-1151 (42309)
FAX: +81 45-566-1587

**Abstract**

To find good sheath materials that react minimally with the superconducting core of iron-based superconducting wires, we investigated the reaction between polycrystalline $SmFeAsO_{0.92}F_{0.08}$ and the following seven metals: Cu, Fe, Ni, Ta, Nb, Cr and Ti. Each of seven metals was prepared as a sheath-material candidate. The interfacial microstructures of $SmFeAsO_{0.92}F_{0.08}$ and these metal-sheath samples were analysed by electron probe micro analyzer after annealing at 1000 °C for 20 h. Amongst the seven metal-sheath samples, we found that Cu was the best, because it reacted only very weakly with polycrystalline $SmFeAsO_{0.92}F_{0.08}$. Moreover, Cu is essential for superconducting wires as a stabilizing material. Metal sheaths made of Fe and Ni do not give rise to reaction layers, but large interdiffusion between these metals and polycrystalline $SmFeAsO_{0.92}F_{0.08}$ occurs. In contrast, Metal sheaths made of Ta, Nb, Cr and Ti do form reaction layers. Their reaction layers apparently prevent electric current from flowing from the sheath material to the superconducting core. In general, Through this research, Cu will be expected not only as a stabilizing material but also as a sheath material for superconducting Sm-1111 wire fabricated by *ex situ* PIT method.

Keywords: Superconducting wire, interface diffusion, Sheath material, EPMA, SmFeAsO

## 1. Introduction

Since the first discovery of layered iron-based LaFeAs(O,F) [1], a series of iron-based superconductors has been reported [2–5]. For iron-based superconductor F-doped SmFeAsO (Sm-1111), a superconducting transition temperature ($T_c$) of 55 K was attained [6]. Because of this high $T_c$, many applications have been attempted, such as superconducting thin films [7–9], Josephson junctions [10,11] and superconducting wires [12–14].

Superconducting wires made from iron-based superconductors may be fabricated by the powder-in-tube (PIT) method [15–17]. Recently, Ma *et al.* fabricated F-doped Sm-1111 wires by the *in situ* PIT method using a silver sheath [13]. The use of a silver sheath led to the first successful *in situ* fabrication of F-doped Sm-1111 wires. The *ex situ* fabrication of F-doped Sm-1111 wire was first reported in our previous study [18]. A disadvantage of *ex situ* fabrication of F-doped Sm-1111 wire is that the F content in $SmFeAsO_{1-x}F_x$ decreases during the heat treatment after rolling and/or wire drawing. This decrease results in the degradation of superconductivity but it is prevented by using a binder composed of $SmF_3$, an arsenide of Fe and an arsenide of Sm. In this previous study, we demonstrated the *ex situ* fabrication of an F-doped Sm-1111 wire with an Ag sheath, and the transport critical current density ($J_c$) attained ~4000 A/cm$^2$ at 4.2 K [18]. In addition, we ascertained that, with this *ex situ* fabrication technique, the silver does not react with the superconducting core during heat treatment. Studies using the boss fabrication processes of F-doped Sm-1111 wire also showed that silver is very appropriate for sheath material because it does not react with the superconducting core during heat treatment [13, 14, 18]. Unfortunately, silver is prohibitively expensive for practical utility in superconducting wires.

A considerable difference exists between *in situ* and *ex situ* fabrication processes. The former employs powder starting materials, whereas the latter employs a powder of synthesized superconducting material. In both processes, these powders are packed into a metal tube and the tubes are annealed after wire drawing. In general, starting materials are more reactive than materials synthesized at a high temperature. Although silver is also appropriate for sheath material of ex situ superconducting wire, other metals are not examined yet. Therefore, for *ex situ* fabrication, we have investigated the reaction between synthesized F-doped Sm-1111 and sheaths composed of various metals to find sheath materials that can substitute for silver.

## 2. Experiment

Polycrystalline $SmFeAsO_{0.92}F_{0.08}$ was synthesized by a solid-state reaction [19]. Samples of Cu, Fe, Ni, Ta, Nb, Cr and Ti were prepared as sheath-material candidates for *ex situ* fabrication. A rectangular parallelepiped of $SmFeAsO_{0.92}F_{0.08}$ was enveloped in the powder of a given metal and pressed into a pellet, following which the pellet was sealed in a quartz tube and annealed at 1000 °C for 20 h. This process was repeated for each metal-sheath sample. Next, these pellets were cut by a diamond cutter and their surfaces were polished. These samples are called structure-enveloped-by-metal samples (SEMS). A schematic of SEMS is shown in figure 1. SEMSs can be divided into three areas: metallic area, Sm-1111 area and reaction-layer area, as shown in figure 1 (b) and (c).

Polycrystalline $SmFeAsO_{0.92}F_{0.08}$ was characterized by X-ray diffraction (XRD; Bruker AXS D8 ADVANCE) using Cu Kα radiation. The electrical resistivity was measured by the standard four-probe technique using Au electrodes at temperatures between 30 and 300 K. Reactions between $SmFeAsO_{0.92}F_{0.08}$ and the metal-sheath samples were analysed by an electron probe micro analyzer (EPMA; SHIMADZU EPMA-8705). Amounts of Fe, Ni, Cr, Cu and Ti were quantified by the intensity of the Kα radiation. Similarly, Amounts of Sm, As, Nb and Ta were quantified by the intensity of the Lα radiation. Figures 1(b) and 1(c) show the locations where EPMA mappings, line scans and quantitative analyses were performed. EPMA mappings were performed in 0.09 or 0.36 $mm^2$ areas that included the border between $SmFeAsO_{0.92}F_{0.08}$ and each metal-sheath sample. Line scan analyses were performed along a straight line across the border. Moreover, quantitative analyses using the EPMA were performed in a 10 $\mu m^2$ area in each of the six following places: the border of the metallic area, the edge of SEMS, the border of the Sm-1111 area, the centre of the Sm-1111 area, the reaction-layer area and a part of the standard sample of polycrystalline $SmFeAsO_{0.92}F_{0.08}$. Provided that the edge of each SEMS is not affected by the diffusional reaction between $SmFeAsO_{0.92}F_{0.08}$ and the metal-sheath sample, the characteristic X-ray intensity of each metal-sheath sample from the border of the metallic area and reaction-layer area is normalized by the characteristic X-ray intensity from the edge of the corresponding SEMS. Similarly, the characteristic X-ray intensity of Sm, Fe and As from the border and the centre of Sm-1111 area are normalized by their characteristic X-ray intensity of a standard sample of $SmFeAsO_{0.92}F_{0.08}$.

## 3. Results and Discussion

*3.1 Characterization of as-grown sample*
Figure 2 shows the XRD pattern of an as-grown sample that mainly consists of the $SmFeAsO_{0.92}F_{0.08}$ phase. Several weak peaks from SmOF, SmAs and FeAs are also detected. Figure 3 shows the temperature dependence of the electrical resistivity $\rho$ for $SmFeAsO_{0.92}F_{0.08}$, which exhibits metallic behaviour at temperatures > 54 K. At $T_c \approx 54$ K, a sharp drop in $\rho$ is observed, which is attributed to superconducting transition.

*3.2 EPMA analysis for each SEMS*
Figure 4 shows EPMA area maps of each SEMS. For Cu, Fe and Ni SEMS, we do not observe reaction layers, whereas for Ta, Nb, Cr and Ti SEMS, reaction layers form along the borders. A previous study did not observe transport critical current in F-doped Sm-1111 wires fabricated by *in situ* PIT with Ta, Nb and Fe/Ti sheaths [20], because their reaction layers apparently prevented electric current from flowing from the sheath material to the superconducting core. For *ex situ* fabrication, such reaction layers form when using Ta, Nb, Cr and Ti. Therefore, the same result as *in situ* fabrication is to be expected. If the interdiffusion between Sm-1111 area and metallic area is sufficiently small, the elemental metal of each metal-sheath sample should be detected only inside the metallic area, and similarly, the elements of Sm, Fe and As should be detected only inside the Sm-1111 area. For Cu SEMS, we do not observe any diffusion of Sm, Fe and As into metallic area. For Fe and Ni SEMS, As is detected in the metallic area as well as the Sm-1111 area because As reacts easily with Fe and Ni. Moreover, a diffusional reaction occurs over a broad range for both elemental Fe and Ni in Ni SEMS.

Figure 5 shows EPMA line scans for each SEMS. Elemental Sm, Fe and As, which are detected in the Sm-1111 area, disappear almost completely at the border of the metallic area for Cu SEMS, whereas elemental Cu is detected locally in the Sm-1111 area of this sample. This suggests that, although Cu barely reacts with $SmFeAsO_{0.92}F_{0.08}$, it interfuses into the grain boundaries of $SmFeAsO_{0.92}F_{0.08}$ during heat treatment. For Ta, Nb, Cr and Ti SEMSs, reaction layers containing Sm, Fe and As were generated with widths of approximately 35, 34, 77 and 79 μm, respectively. In the metallic area of Ta, Nb, Cr and Ti SEMS, no elemental Sm, Fe or As was detected. Among SEMSs with reaction layers, only Cr SEMS exhibits diffusional reaction of Cr into the Sm-1111 area, and in Ni SEMS, Ni was detected beyond the border. Therefore, Ni and Cr diffuse easily into the Sm-1111 area. In Fe SEMS, the metallic area contains more As element than the Sm-1111 area. This result suggests that As is more stable in the metallic area than in the Sm-1111 area.

Figure 6 shows the quantitative analyses using the EPMA in the Sm-1111 area. This figure shows the compositional change from stoichiometric $SmFeAsO_{0.92}F_{0.08}$, which is induced by diffusion and reaction between $SmFeAsO_{0.92}F_{0.08}$ and each metal-sheath sample during a heat treatment. As normalized intensities approach unity, the changes in elemental composition of Sm-1111 area become smaller. For Cu, Ta and Nb SEMSs, the elemental composition of $SmFeAsO_{0.92}F_{0.08}$ are almost same as standard $SmFeAsO_{0.92}F_{0.08}$. For Cr and Ti SEMSs, although the some elements remain unchanged stoichiometrically, Fe in Cr SEMS and As in Ti SEMS diffuse into reaction-layer area. For Fe and Ni SEMSs, significant compositional changes from stoichiometric $SmFeAsO_{0.92}F_{0.08}$ are observed.

Figure 7 shows the quantitative analyses using the EPMA in the metallic area and reaction-layer area. This figure exhibits the change of metal purity near the border of the metal area and near the reaction-layer area, which is induced by interdiffusion between $SmFeAsO_{0.92}F_{0.08}$ and each metal-sheath sample during the heat treatment. When the normalized intensity approaches unity, the metal purity in measured places becomes higher. The metal purity of Cu near the border is almost same as the edge of Cu SEMS. For Ta, Nb, Cr and Ti SEMSs, degradation of metal purity are small outside the reaction layers, and are on the order of the observed metal purity for Cu SEMS.

These results indicate Cu is the best sheath material in the seven metal-sheath samples. Moreover, In fabricating the superconducting wire, Cu is essential as a stabilizing material. The role of this Cu is to minimize Joule heating in case of breaking of superconducting phase. Therefore, using Cu will be expected with two effects as a sheath material and a stabilizing material.

## 4. Conclusion

To find good sheath materials for iron-based superconducting Sm-1111 wires, we prepared seven metal-sheath samples. The advantage of the SEMSs approach is that various metals can be investigated with only a small amount of superconducting powder. In this study, we categorized these seven metal-sheath samples into four groups: The first contains only Cu, which does not form a reaction layer (indeed, it barely reacts at all with the Sm-1111 phase), although a little interfusion appears in the Sm-1111 area. The second contains Fe and Ni, which also do not form reaction layers, but do display a large amount of interdiffusion between $SmFeAsO_{0.92}F_{0.08}$ and metals of Fe and Ni. The third contains Ta and Nb, which form reaction layers, but the elemental composition of the Sm-1111 area remains unchanged. The last contains Cr and Ti, which also form reaction layers and cause small changes in the composition of the Sm-1111 area. Therefore, we conclude that, from among these seven metals, Cu is the best sheath material. Moreover, Cu is essential as a stabilizing material for superconducting wires. Therefore, Cu will be expected as not only a stabilizing material but also a sheath material for superconducting Sm-1111 wire fabricated by *ex situ* PIT technique.

**Acknowledgments**

This work was partially supported by the Research Grant of Keio Leading-edge Laboratory of Science & Technology

Figure 1. Schematic of SEMSs. (a) Samples before annealing. (b) Cu, Fe and Ni SEMSs. These SEMSs do not have reaction layers. (c) Ta, Nb, Cr and Ti SEMSs. These SEMSs do have reaction layers. The black and grey regions correspond to polycrystalline $SmFeAsO_{0.92}F_{0.08}$ and metal-sheath sample, respectively. The blue square and the green line represent the areas of the EPMA mapping and line scan, respectively. The small red-X labels represent regions where quantitative analysis was performed. The large red-X labels at the edge of SEMS represent regions where characteristic X-ray intensity is measured to perform the normalization.

Figure 2. Powder XRD pattern of bulk $SmFeAsO_{0.92}F_{0.08}$. Red bars at the bottom show the calculated Bragg diffraction positions of $SmFeAsO_{0.92}F_{0.08}$. The red circle, blue triangle and green square denote peaks of SmOF, SmAs and FeAs, respectively.

Figure 3. Resistivity $\rho$ versus temperature $T$ for bulk $SmFeAsO_{0.92}F_{0.08}$.

Figure 4. EPMA-area mapping for each SEMS. The rows correspond to a given sample. The first column shows (scanning electron microscope) SEM images; the second through fifth columns show the EPMA area mappings of metals, Sm, Fe and As, respectively.

Figure 5. EPMA line scans for each SEMS. Blue downward triangles show the Kα or Lα emission spectrum for each sample. Also shown are the Sm Lα emission spectrum (black squares), the Fe Kα emission spectrum (red circles) and the As Lα emission spectrum (green upward triangles).

The origin of the abscissa is defined as the position at which the Sm intensity approaches zero and the ordinate is normalized by the average of the Sm intensity near the centre of $SmFeAsO_{0.92}F_{0.08}$.

Figure 6. Quantitative analysis of the centre and border of Sm-1111 area for SEMSs. characteristic X-ray intensities of Sm, Fe and As are normalized by their characteristic X-ray intensities for the standard $SmFeAsO_{0.92}F_{0.08}$. Open and solid squares represent measurements in the border and centre, respectively. Black squares, red circles and blue triangles represent measurements of Sm, Fe and As, respectively.

Figure 7. Quantitative analysis of the border of metallic area for SEMSs. characteristic X-ray intensities of metallic and reaction-layer area are normalized by the characteristic X-ray intensity of the edge of each SEMS. Open diamonds represent measurements at the border of the metallic area and solid diamonds represent measurements inside the reaction-layer area.

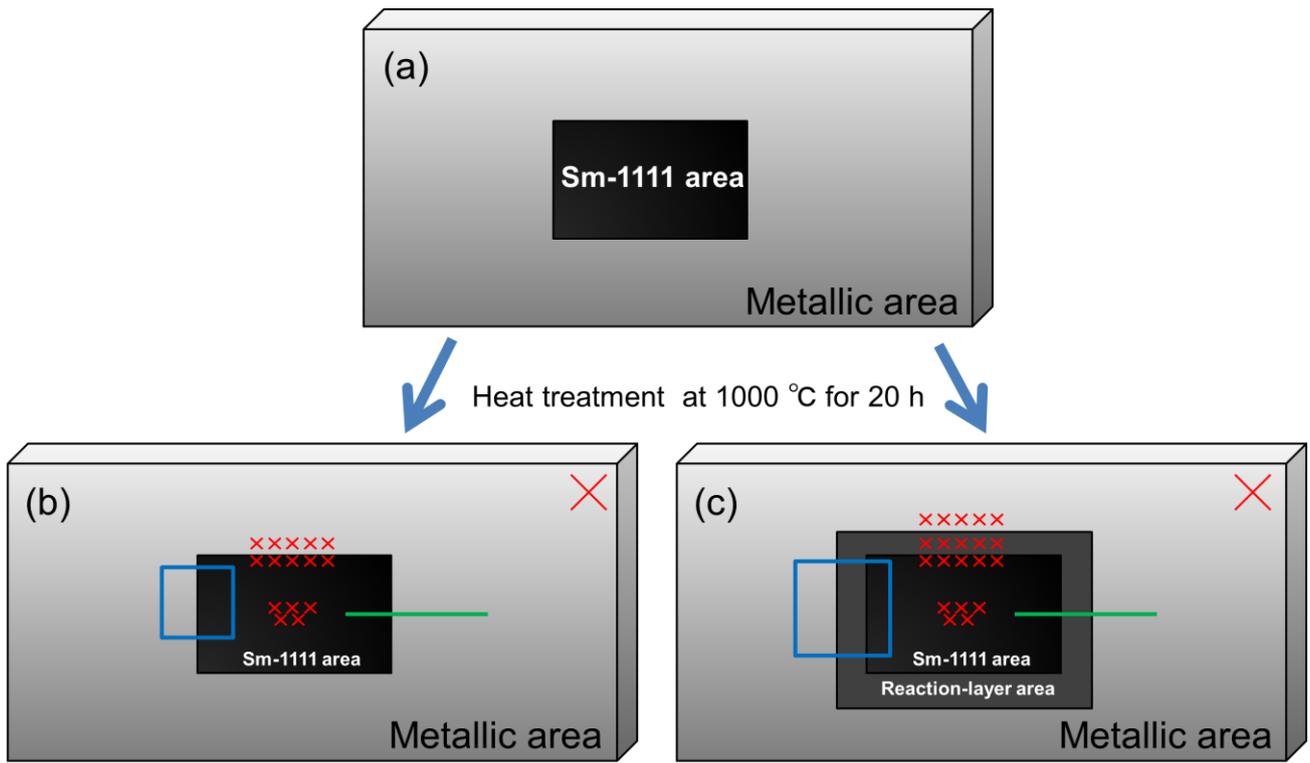

Fig. 1.

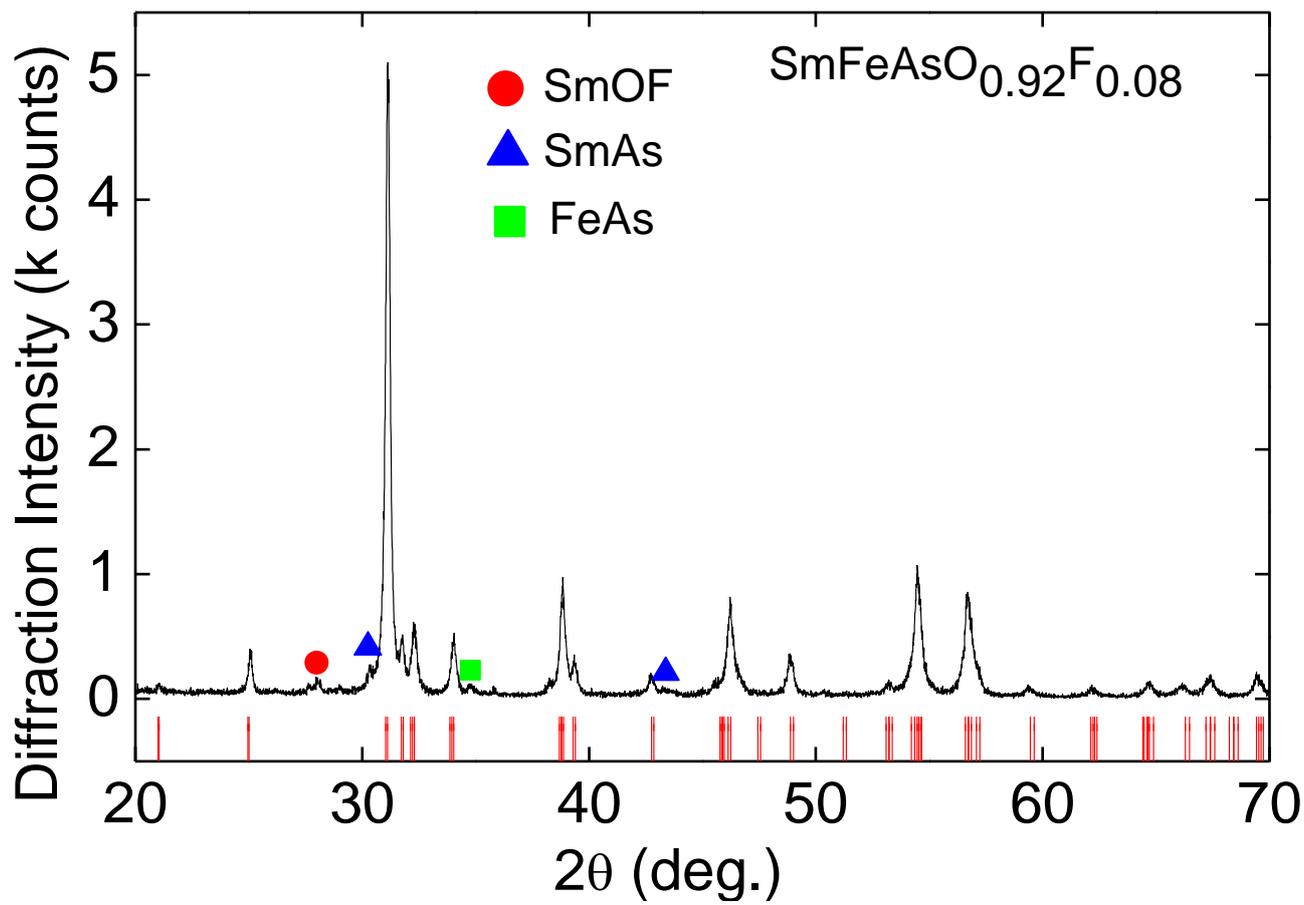

Fig. 2.

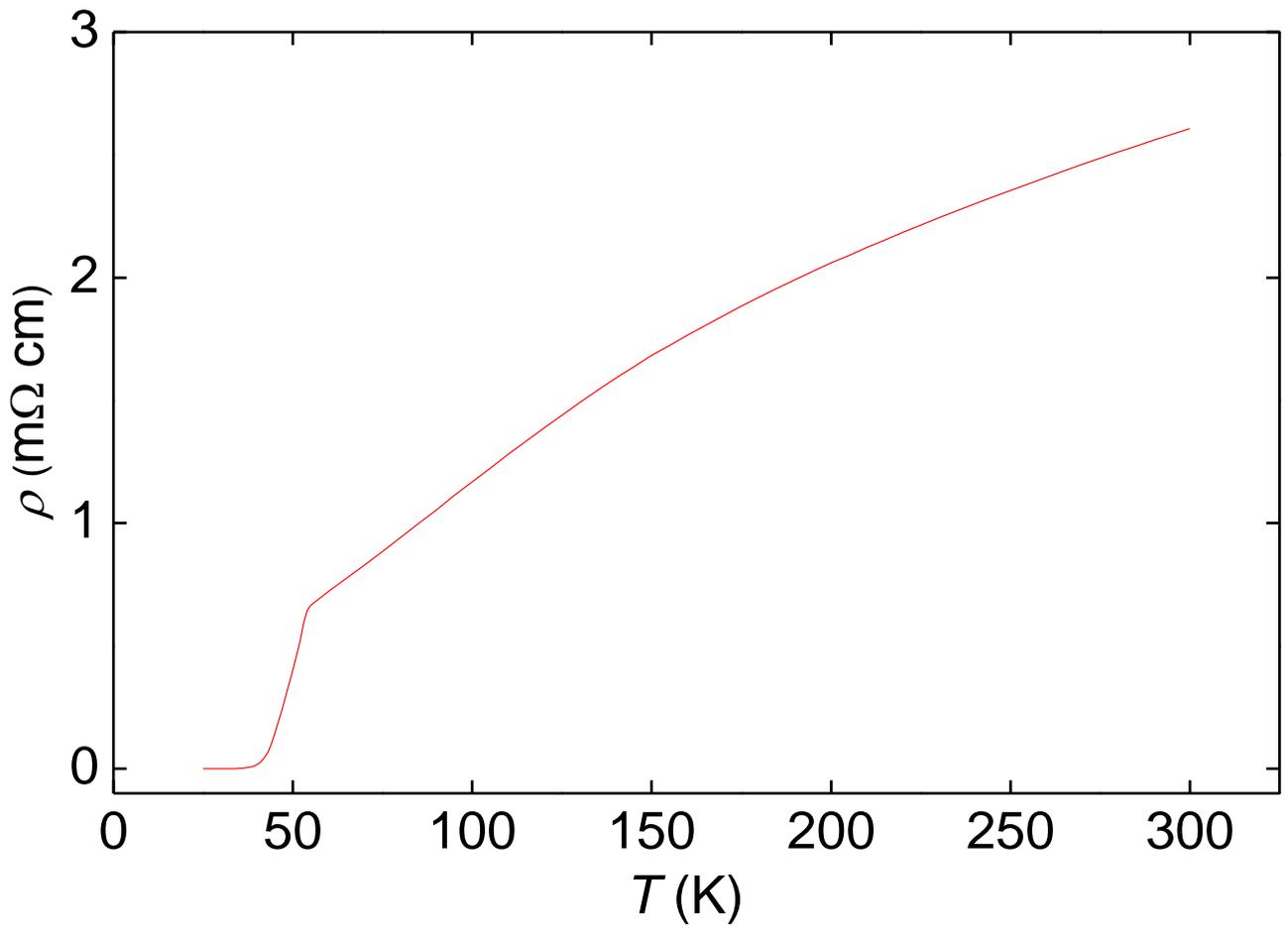

Fig. 3.

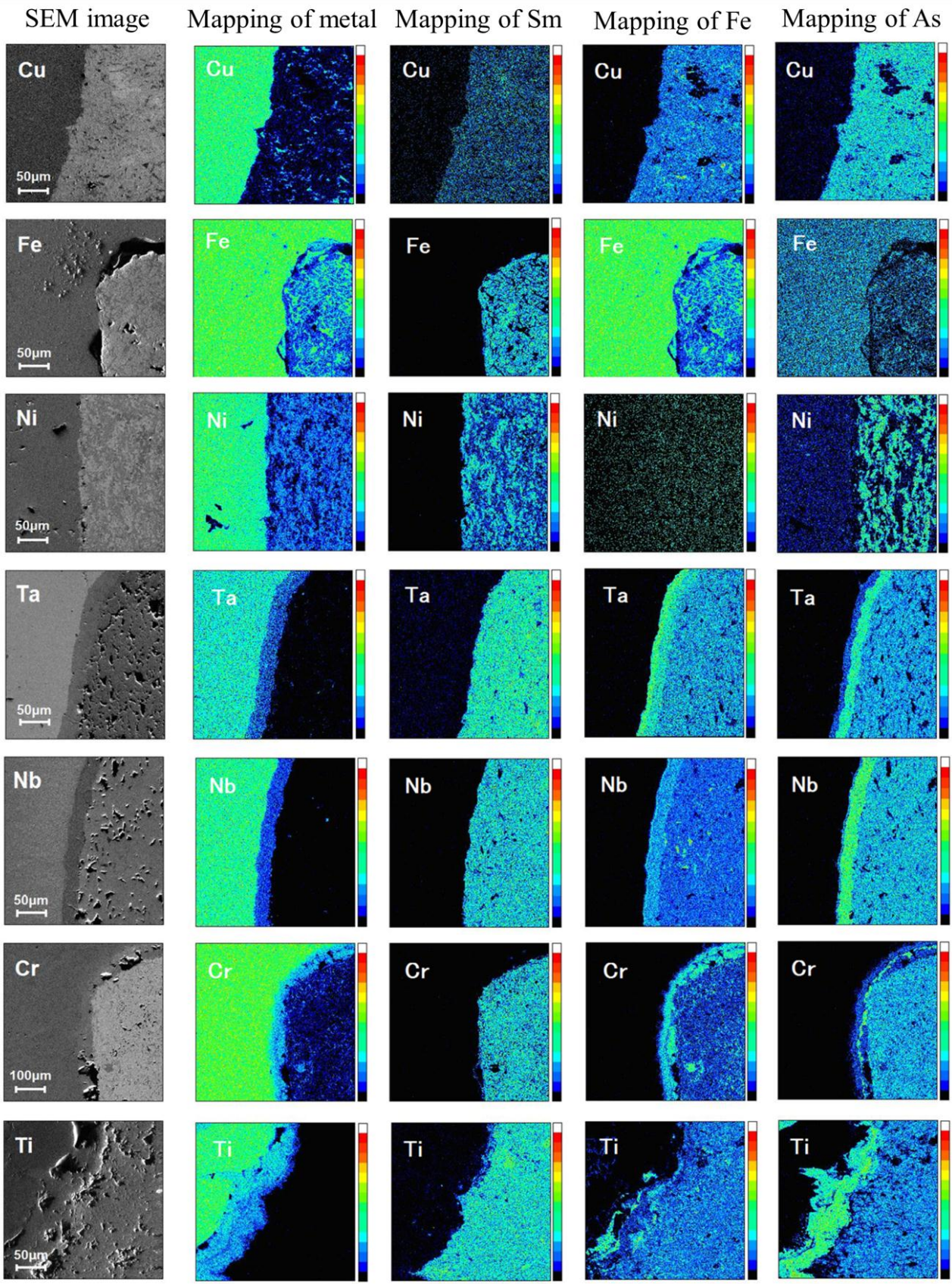

Fig. 4.

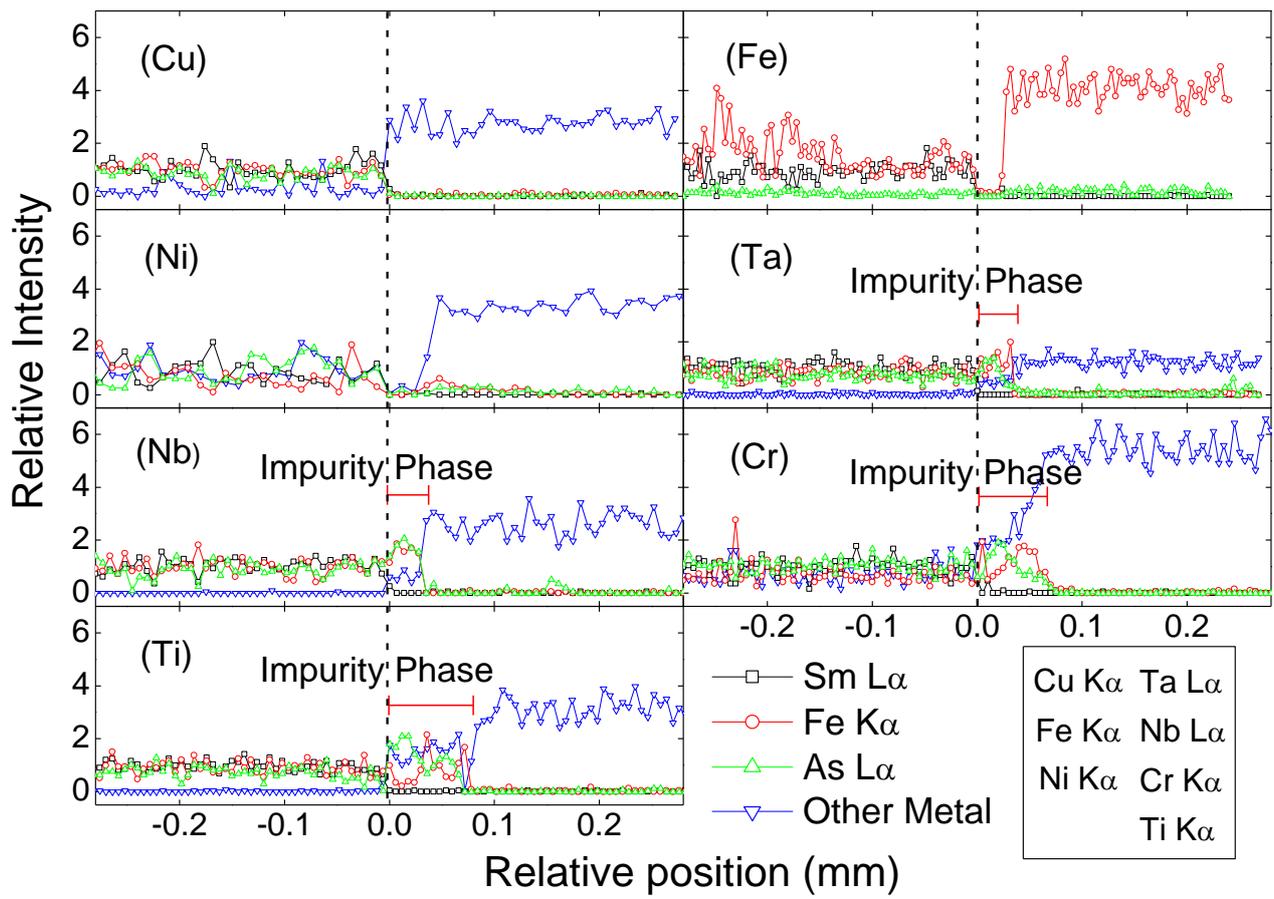

Fig. 5.

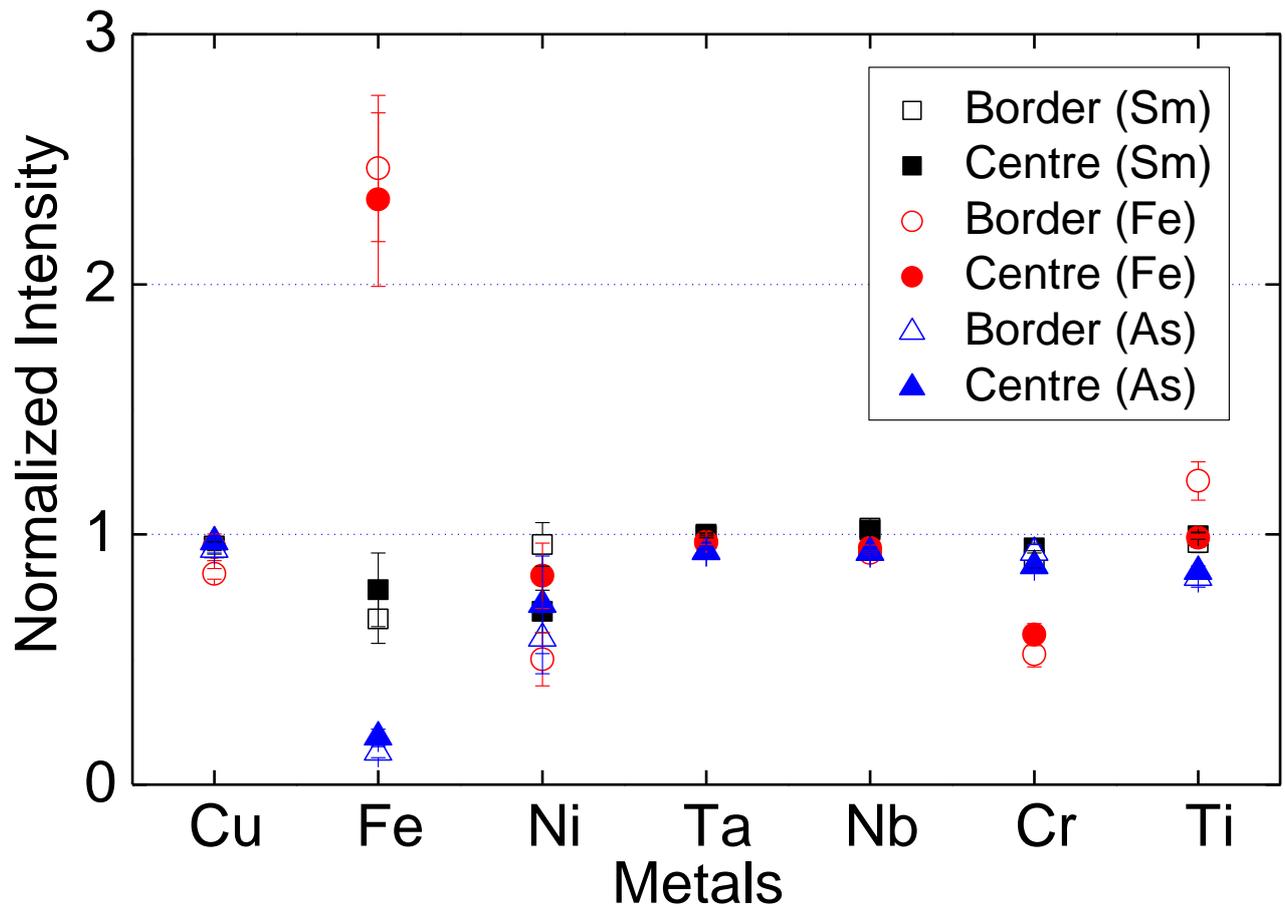

Fig. 6.

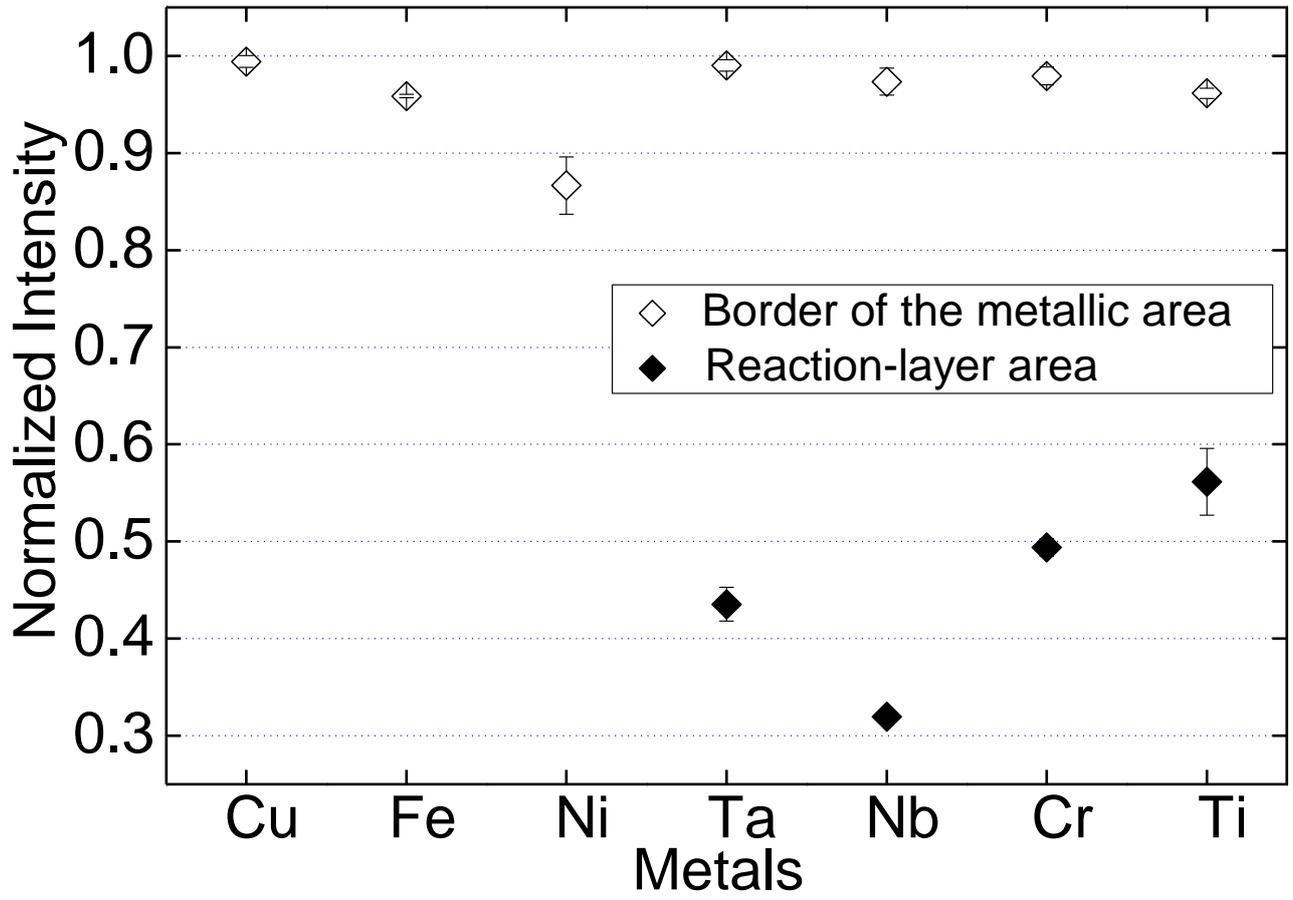

Fig. 7.